\documentclass[a4paper,11pt]{article}
\pdfoutput=1 
\usepackage{jheppub} 

\usepackage[T1]{fontenc} 

\RequirePackage[displaymath]{lineno} 
\usepackage{epstopdf}
\usepackage{epsfig}
\usepackage{graphicx}
\usepackage{dcolumn}
\usepackage{bm}
\usepackage{ltablex,booktabs}
\usepackage{overpic}
\usepackage{subfigure}
\usepackage{float}
\usepackage{color}
\usepackage{amsmath}
\usepackage{mathcomp}
\usepackage{mathrsfs}
\usepackage{multirow}
\usepackage{rotating}
\usepackage{amssymb}
\usepackage{gensymb}
\usepackage{amsmath}
\usepackage{tabularx}
\usepackage{threeparttable}
\usepackage{graphicx}

\title{Measurement of Born Cross Sections and Effective Form Factors of \boldmath{$e^+e^-\to \Omega^{-}\bar{\Omega}^{+}$} from \boldmath{$\sqrt{s}$ = 3.7 to 4.7~GeV}}

\collaboration{BESIII Collaboration}
\emailAdd{besiii-publications@ihep.ac.cn}

\abstract{Using $e^+e^-$ collision data corresponding to an integrated luminosity of 22.7 fb$^{-1}$, collected at center-of-mass energies between 3.7 and 4.7 GeV with the BESIII detector at the BEPCII storage ring, we measure the energy-dependent Born cross sections of $e^+e^-\to \Omega^{-}\bar{\Omega}^+$ and the effective form factors of the $\Omega^-$ baryon. The analysis employs a single baryon tagging method, and the results are consistent with theoretical predictions, providing critical constraints on the electromagnetic structure of the $\Omega^-$ hyperon. No significant signal of charmonium or charmonium-like states decaying to $\Omega^{-}\bar{\Omega}^+$ is observed in the investigated energy range.}

\arxivnumber{}

\begin{document}
\maketitle
\flushbottom

\section{Introduction}

Over the past two decades, many vector states have been observed at
center-of-mass (c.m.) energies between 3.7 and 4.7 GeV by various
$e^+e^-$ colliders. Four charmonium states ($\psi(3770)$,
$\psi(4040)$, $\psi(4160)$, and $\psi(4415)$) have been observed in
the inclusive hadronic cross section line shape~\cite{pdg, BES:2007zwq}. Several charmonium-like vector states, such as
$Y(4220)$, $Y(4260)$, $Y(4360)$, and $Y(4660)$, have been observed by
the BaBar~\cite{PRL95-142001, PRL98-212001, PRD86-051102,
  PRD89-111103}, Belle~\cite{PRL99-142002, PRL99-182004,
  PRL110-252002, PRD91-112007}, CLEO~\cite{PRL96-162003,
  PRD74-091104}, and BESIII~\cite{CPC38-043001, PRL114-092003,
  PRD91-112005, PRL118-092001, PRD104-052012} experiments. Many
theoretical models have been proposed to understand the underlying
structure of these $Y$ states, interpreting them as hybrid
charmonia~\cite{hybrid}, tetraquarks~\cite{tetraquark}, or hadronic
molecules~\cite{molecule}. However, none of them has been able to
describe all the experimental observations. Up to now, no evidence of
their decays into light-quark baryon-antibaryon pairs has been
found. As an example, BESIII recently searched for charmonium-like
vector states in the processes $e^+e^-\to \Xi^-\bar{\Xi}^{+}$ and
$e^+e^-\to \Lambda\bar{\Lambda}$~\cite{PRD104-L091104,
  PRL124-032002}. In addition, BESIII reported a large number of
precision measurements of the cross sections of SU(3)
baryon-antibaryon pairs~\cite{arXiv:2111.08425} and some experimental
results regarding the production of $\Omega^{-}\bar{\Omega}^{+}$ pairs~\cite{pdg}.

The electromagnetic structure of the $\Omega^-$ hyperon, consisting of
three valence strange quarks, remains poorly understood due to limited
experimental measurements. As a baryon with a spin-parity of $3/2^+$,
its structure is characterized by its electric charge
($G_{\text{E0}}$), magnetic dipole ($G_{\text{M1}}$), electric
quadrupole ($G_{\text{E2}}$), and magnetic octupole ($G_{\text{M3}}$)
form factors~\cite{Nozawa:1990gt, Korner:1976hv}. Precision
measurements of the $\Omega^-$ electromagnetic form factors in the
space-like region ($q^2 < 0$) are challenging due to the difficulty in
creating suitable strange baryon targets which can be
scattered by electron beams. Currently, the primary method of measuring
the electromagnetic form factors of hadrons is through the cross
section measurement in $e^+e^-$ annihilation, which offers information
of time-like momentum transfer ($q^2 > 0$).  Due to limited
statistical precision, most experiments have only been able to measure
a combination of the electromagnetic form factors, known as the
effective form factor (EFF)~\cite{PRL124-032002}. The form factors of $SU(3)$ baryons can
be obtained from the production cross sections of $e^+e^-\to
\gamma^{*}\to B\bar{B}$. Recent measurements reported the $\Omega^-$
electromagnetic form factors at a c.m. energy ($\sqrt{s}$) of
approximately 3.7 GeV ($q^2 = s = 14.2~$GeV$^2$) using the
data from the BESIII and CLEO-c
detectors~\cite{PRD-052003,PRD96-092004,PLB739-90}. These data provide
valuable information about the time-like form factors in the large
$q^2$ region and offer a unique opportunity to test their behaviour
under extreme conditions. A recent calculation of the EFF for the
$\Omega^-$ hyperon was performed using the covariant spectator quark
model~\cite{PRD-074018}. However, the
accuracy of the description of the EFF is limited due to insufficient
data. More data at larger momentum transfer are therefore highly
desirable. The measurements would provide better constraints on the
shape of the form factors at large $q^2$ and enhance the
significance of time-like measurements in calibrating different
theoretical models.

In this article, we present a measurement of the Born cross sections
(BCSs) of $e^+e^-\to \Omega^{-}\bar{\Omega}^+$ and the EFFs for $\Omega^-$ using data corresponding to a total
integrated luminosity of 22.7 fb$^{-1}$, collected at c.m. energies
between 3.7 and 4.7 GeV with the BESIII detector~\cite{Ablikim:2009aa}
at the Beijing Electron Positron
Collider~(BEPCII)~\cite{Yu:IPAC2016-TUYA01}. Through the analysis of
the cross-section line shape measured by the single baryon tag method,
we can search for charmonium-like vector states in their baryonic
decay modes. In addition, the EFFs obtained will provide essential
information to understand the structure of $\Omega^-$ hyperons.

\section{Detector and data sets}
\label{sec:detector_dataset}


The BESIII detector~\cite{Ablikim:2009aa} records symmetric $e^+e^-$
collisions provided by the BEPCII storage
ring~\cite{Yu:IPAC2016-TUYA01} in the c.m. energy range from
2.0 to 4.95~GeV, with a peak luminosity of $1 \times
10^{33}\;\text{cm}^{-2}\text{s}^{-1}$ achieved at $\sqrt{s} =
3.77\;\text{GeV}$.  BESIII has collected large data samples in this
energy region~\cite{Ablikim:2019hff, EcmsMea, EventFilter}.

Its cylindrical core consists of a helium-based multilayer drift
chamber~(MDC), a plastic scintillator time-of-flight system~(TOF), and
a CsI~(Tl) electromagnetic calorimeter~(EMC), which are all enclosed
in a superconducting solenoidal magnet that provides a 1.0~T magnetic
field. The solenoid is supported by an octagonal flux-return yoke with
resistive plate counter muon identifier modules interleaved with
steel. The acceptance of charged particles and photons is 93\% over
$4\pi$ solid angle. The charged-particle momentum resolution at
1.0~GeV/$c$ is $0.5\%$, and the specific ionization energy
loss~(d$E$/d$x$) resolution is $6\%$ for electrons from Bhabha
scattering. The EMC measures photon energies with a resolution of
$2.5\%$~($5\%$) at $1$~GeV in the barrel (end cap) region. The time
resolution in the TOF barrel region is 68~ps, while that in the end
cap region was 110~ps. The end cap TOF system was upgraded in 2015
using multigap resistive plate chamber technology, providing a time
resolution of 60~ps, which benefits 66.6\% of the data used in this
analysis~\cite{tof1, tof2, etof}.

GEANT4-based~\cite{GEANT4-Col,GEANT4-Allison} Monte Carlo (MC)
simulations are used to evaluate the detection efficiency and estimate
the physics backgrounds. The simulations include the geometric and
material description of the BESIII detectors~\cite{Huang:2022wuo}, the
detector response and digitization models, as well as the tracking of
the detector's running conditions and performance. The simulations
model the beam energy spread and initial state radiation (ISR) in the
$e^+e^-$ annihilations with the {\sc kkmc} generator~\cite{kkmc}. The
two signal processes, $e^+e^-\to \Omega^{-}\bar{\Omega}^+$,
$\Omega^-\to\Lambda K^-$, $\Lambda\to p\pi^-$, $\bar{\Omega}^+\to$
anything and $e^+e^-\to \Omega^{-}\bar{\Omega}^+$, $\bar{\Omega}^+\to
\bar{\Lambda}K^+$, $\bar{\Lambda}\to \bar{p}\pi^{+}$, $\Omega^-\to$
anything, are generated to evaluate the detection efficiency. For each signal
process, 100,000 events are simulated with the joint angular
distribution of the decay sequence from Ref.~\cite{songjiaojiao} at
each of the thirty-four c.m. energies.  The decay of the
$\Omega^-$ hyperon are generated by {\sc
  evtgen}~\cite{evtgen,simulation-RGPing}, with branching fractions
set to the world average values from the Particle Data Group
(PDG)~\cite{pdg}. In addition, large simulated generic MC samples
are generated for background study~\cite{topo}.

\section{Event selection and signal determination}
\label{chap:event_selection}

 A single baryon tag method is used to select the $e^+e^-\to
 \Omega^-\bar{\Omega}^{+}$ candidate events, i.e. only one baryon of
 the $\Omega^-\bar{\Omega}^{+}$ pair is reconstructed in each event,
 where the $\Omega^-$ or $\bar{\Omega}^{+}$ is reconstructed with
 $\Omega^-\to K^-\Lambda$, $\Lambda\to p\pi^-$ or $\bar{\Omega}^+\to
 K^+\bar{\Lambda}$, $\bar{\Lambda}\to \bar{p}\pi^{+}$, and the
 $\Omega^-$ or $\bar{\Omega}^{+}$ on the recoil side is inferred from
 the recoil mass of the reconstructed particles. The following event
 selections are described for $\Omega^-\to K^-\Lambda$ ($\Lambda\to p\pi^-$), as an example;
 the same selections are also applied for the $\bar{\Omega}^{+}$
 selection.
  
All charged tracks detected in the MDC must satisfy
$|\rm{cos\theta}|<0.93$, where $\theta$ is the polar angle of the
charged track with respect to the $z$-axis, which is the symmetry axis
of the MDC. Particle identification (PID) for charged tracks combines
measurements of the d$E$/d$x$ in the MDC and the flight time in the
TOF to form likelihoods $\mathcal{L}(h)~(h=p,K,\pi)$ for each hadron
$h$ hypothesis. Each charged track is then assigned to the particle
type with the greatest likelihood. Events with at least one charged
proton, kaon, and pion candidate are kept for further analysis.

Each $\Lambda$ candidate is reconstructed with two oppositely charged
$p\pi^-$ tracks, identified using PID algorithm. They are constrained to
originate from a common vertex, and a loose requirement on the
invariant mass of $p\pi^-$ is applied as 1.09 < $M(p\pi^-)$ <
1.14~GeV$/c^{2}$ to suppress background events. The $\Omega^-$
candidates are then reconstructed from all pairs of the selected
$K^-\Lambda$ candidates, which are constrained to originate from a
common vertex. A secondary vertex fit~\cite{vertexfit} is performed to ensure the $\Omega^-$ candidates originate from the interaction point.

To further suppress background, the decay lengths of $\Lambda$ and
$\Omega^-$ candidates, i.e. the distance between the average position
of the $e^+e^-$ collisions and the decay vertex of $\Lambda$ and
$\Omega^-$, are required to be greater than twice the vertex
resolution. The $\chi^2$ values of the vertex fit for $\Lambda$ and
$\Omega^-$ are both required to be less than 100 for further
analysis. If there is more than one $\Omega^-$ or $\bar{\Omega}^{+}$
candidate satisfying all above requirements in the event, the one with
the minimum $\chi^2$ is retained for further study.

\begin{figure}[hbtp]
\begin{center}
\includegraphics[width=.55\textwidth]{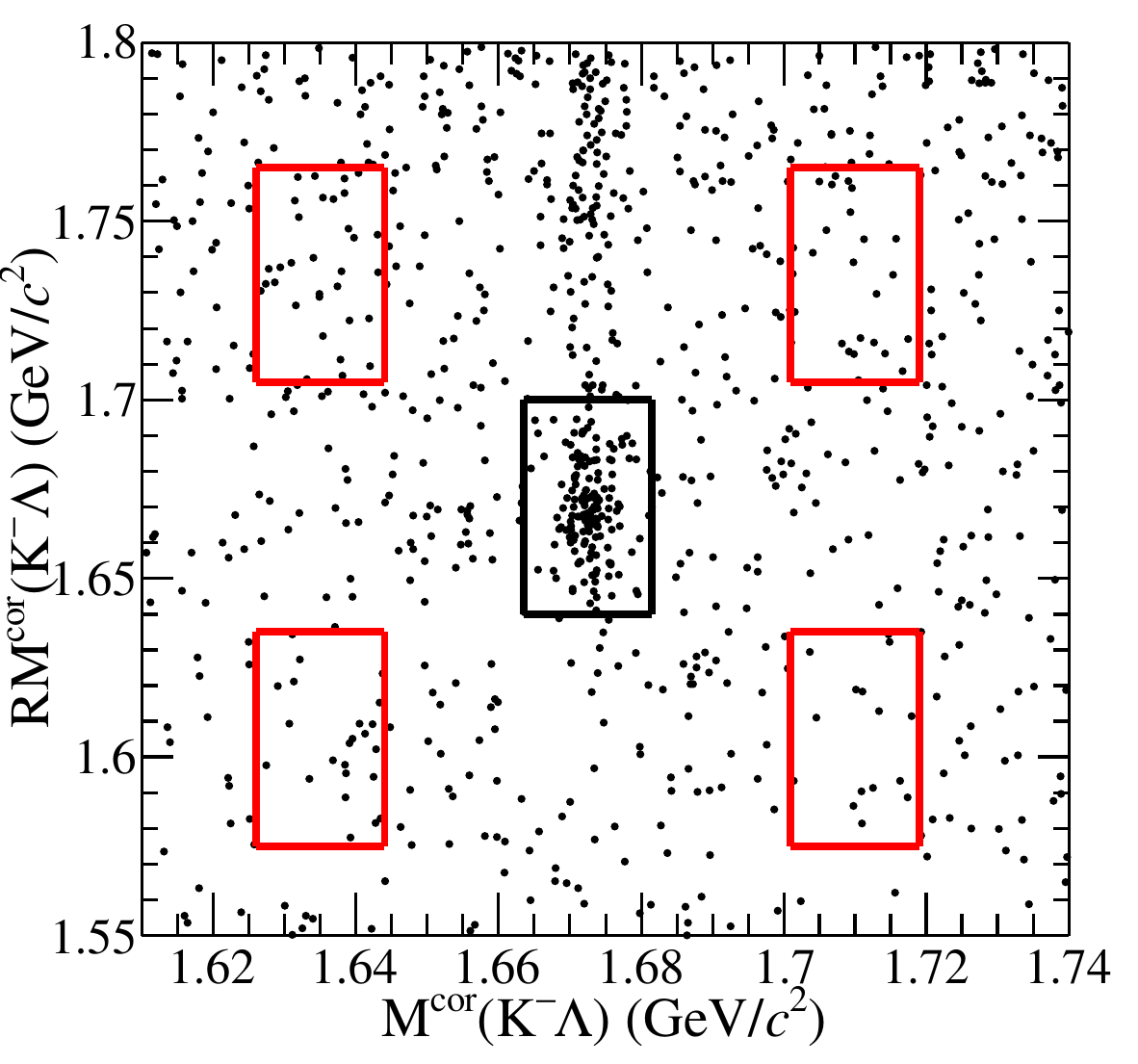}
\caption{Distribution of $RM^{\rm cor}(K^-\Lambda)$ versus $M^{\rm cor}(K^-\Lambda)$ of the accepted candidates from all thirty-four
  c.m. energies, where the black box shows the signal region, and the
  red boxes denote the selected sideband regions.}
\label{fig_2d_omega_romega}
\end{center}
\end{figure}

To improve the mass resolution, we exploit the correlation between
$M(K^-\Lambda)$ and $M(p\pi^-)$ by using the variable $M^{\rm
  cor}(K^-\Lambda) = M(K^-\Lambda) - M(p\pi^-) + m(\Lambda)$ instead
of $M(K^-\Lambda)$, where $M(K^-\Lambda)$ is the invariant mass of the
$K^-\Lambda$ combination and $m(\Lambda)$ is the known mass of the
$\Lambda$ from the PDG~\cite{pdg}. Similarly, for the recoil mass, we
use $RM^{\rm cor}(K^-\Lambda) = RM(K^-\Lambda) + M(K^-\Lambda) -
m(\Omega^-)$ instead of $RM(K^-\Lambda$), where $m(\Omega^-)$ is the
known mass of the $\Omega^-$ from the
PDG~\cite{pdg}. Figure~\ref{fig_2d_omega_romega} shows the
distribution of $RM^{\rm cor}(K^-\Lambda)$ versus $M^{\rm cor}(K^-\Lambda)$ for all data samples having passed all above
requirements and with a requirement of $M(p\pi^-)$ in the $\Lambda$
signal region of [1.111, 1.121]~GeV$/c^{2}$, corresponding to three
times the $p\pi^-$ mass resolution to the nominal $\Lambda$
mass~\cite{pdg}. A clear enhancement around the nominal $\Omega^-$
mass is observed.

After applying the above criteria, the survived background events are
mainly from non-$\Omega$ processes, such as $e^+e^-\to \Lambda
\bar{\Lambda} \phi$ with $\phi \to K^- K^+$. The enhancement above the signal region in $RM^{\rm cor}(K^-\Lambda)$ distribution is due to the process $e^+e^-\to \gamma_{ISR}\psi(3773)$ with $\psi(3773) \to \Omega^{-}\bar{\Omega}^{+}$. The contribution of this background in signal region is negligible according to simulation. The background
contribution in the signal region is estimated by four sideband
regions, each with the same area as the signal region. The
$\Omega^-$ signal and the sideband regions are shown in
Figure~\ref{fig_2d_omega_romega}. The signal region is chosen as
[1.6635, 1.6815]~GeV$/c^{2}$ for $M^{\rm cor}(K^-\Lambda)$ corresponding to
three times the mass resolution in the tag side and [1.64,1.70]~GeV$/c^{2}$ for $RM^{\rm cor}(K^-\Lambda)$ in the recoil side. The sideband regions are chosen as [1.626,
  1.644] and [1.701, 1.719]~GeV$/c^{2}$ for $M^{\rm cor}(K^-\Lambda)$
and [1.575, 1.635]~GeV$/c^{2}$ and [1.705, 1.765]~GeV$/c^{2}$ for
$RM^{\rm cor}(K^-\Lambda)$. The net signal yield ($N^{\rm sig}$) at
each c.m. energy is obtained by subtracting the normalized total number of events in the sideband regions ($\frac{1}{4}N^{\text{bkg}}$) from the number of events in the signal region ($N^{\text{obs}}$) using the formula $N^{\text{sig}} = N^{\text{obs}} - \frac{1}{4}N^{\text{bkg}}$. Table~\ref{cros_list} lists the signal yield at each c.m. energy, where the negative values of $N^{\rm sig}$ have been
set to zero to prevent an unphysical number of signal events. For c.m. energy points where $S(\sigma)$ < 3$\sigma$, we calculate the 90\% confidence level upper limit ($N^{\rm UL}$) using the TRolke package~\cite{TRolke}, after incorporating systematic uncertainties. The signal significance $S(\sigma)$ is defined as $N^{sig}$/$\sqrt{N^{\rm obs}}$ 
and is expressed in units of standard deviations ($\sigma$), which correspond to the probability (p-value) that the observed signal arises from a background fluctuation.

\section{Measurement of BCS and EFF}
\label{chap:BFMEASUREMENT}
The BCS of the $e^+e^-\to \Omega^-\bar{\Omega}^+$ process at each c.m. energy is calculated by
\begin{eqnarray}
\begin{aligned}
\sigma^{\rm {B}}(e^+e^-\to\Omega^-\bar{\Omega}^+)
&= \frac{N^{\rm sig}}{\mathcal{L}(1+\delta)\frac{1}{\left|{1-{\Pi}}\right|^2}\mathcal{B}(\Omega^-\to K^-\Lambda)\mathcal{B}(\Lambda\to p\pi^-)\epsilon},
\label{eq2}
\end{aligned}
\end{eqnarray}
where $N^{\rm sig}$ is the net signal yield, $\mathcal{L}$ is the
integrated luminosity, $\epsilon$ is the detection efficiency,
($1+\delta$) is the ISR correction factor, and (1/$\left|{1-{\Pi}}\right|^2$) is the
vacuum polarization (VP) correction factor~\cite{EPJC-66-585}. The
values of $\mathcal{B}(\Omega^-\to K^-\Lambda)$ =
(67.8$\pm$0.7)\%~and $\mathcal{B}(\Lambda\to p\pi^-)$ =
(63.9$\pm$0.5)\%~are from the PDG \cite{pdg}. An iterative procedure
is adopted to obtain a stable product of
$(1+\delta)\epsilon$~\cite{PRD104-L091104, PRD-052003}. The values
of $(1+\delta)\epsilon$ after the third iteration are
convergent and used to calculate the BCSs.

Assuming that the one-photon exchange $e^+e^-\to \gamma^{*}\to
\Omega^{-}\bar{\Omega}^{+}$ is the dominant process, the EFF |$G_{\rm
  eff}(s)$| for baryons with spin-parity $3/2^+$ is defined by a
combination of the four form factors: $G_{\text{E0}}$,
$G_{\text{M1}}$, $G_{\text{E2}}$, and
$G_{\text{M3}}$~\cite{PRD-052003}. It is written as
\begin{eqnarray}
\begin{aligned}
\lvert G_{\rm eff}(s) \rvert
&= \sqrt{\frac{2\times{\frac{s}{4m^{2}}}\lvert G^{*}_{\rm M}(s)\rvert^{2} + \lvert G^{*}_{\rm E}(s)\rvert^{2}}{2\times{\frac{s}{4m^{2}}+1}}},
\label{eq3}
\end{aligned}
\end{eqnarray}
where {\it{s}} is the square of the c.m. energy, and $m$ is the
$\Omega^-$ baryon mass. $\lvert G^{*}_{\rm E}(s)\rvert$ and $\lvert
G^{*}_{\rm M}(s)\rvert$ are defined as
\begin{eqnarray}
\begin{aligned}
\lvert G^{*}_{\rm E}(s)\rvert^{2}
&=2\lvert G_{\rm E0}\rvert^{2}+{\frac{8}{9}} ({\frac{s}{4m^{2}}})^{2} \lvert G_{\rm E2}\rvert^{2},
\label{eq4}
\end{aligned}
\end{eqnarray}
\begin{eqnarray}
\begin{aligned}
\lvert G^{*}_{\rm M}(s)\rvert^{2}
&={\frac{10}{9}}\lvert G_{\rm M1}\rvert^{2}+{\frac{32}{5}} ({\frac{s}{4m^{2}}})^{2} \lvert G_{\rm M3}\rvert^{2}.
\label{eq5}
\end{aligned}
\end{eqnarray}

The relationship between the BCS and $G_{\rm eff}(s)$ of the $e^+e^-\to \Omega^-\bar{\Omega}^+$ process is defined as

\begin{eqnarray}
\begin{aligned}
\sigma^{\rm {B}}(s)
&={\frac{4\pi\alpha^{2}\mathcal{C}\beta}{3s}} \left [ \lvert G^{*}_{\rm M}(s)\rvert^{2}+{\frac{2m^{2}}{s}} \lvert G^{*}_{\rm E}(s)\rvert^{2}\right ],
\label{eq6}
\end{aligned}
\end{eqnarray}
where $\alpha$ is the fine structure constant, the variable $\beta$ = $\sqrt{1-\frac{4m^{2}}{s}}$ is the velocity of $\Omega^-$, and the Coulomb factor $\mathcal{C}$ parameterizes the electromagnetic interaction between the outgoing baryon and the anti-baryon~\cite{EPJ-3152009}.

$G_{\rm eff}(s)$ is then determined from the BCSs by
\begin{eqnarray}
\begin{aligned}
\lvert G_{\rm eff}(s) \rvert
&= \sqrt{\frac{3s\sigma^{\rm {B}}}{4\pi\alpha^{2}\mathcal{C}\beta(1+\frac{2m^{2}}{s})}}.
\label{eq7}
\end{aligned}
\end{eqnarray}

The measured BCSs and EFFs at the thirty-four c.m. energies are listed in Table~\ref{cros_list}. The upper limits are determined with the profile likelihood method after incorporating the systematic uncertainties~\cite{TRolke}. Figure{~\ref{fig_cros} shows the obtained BCSs at different c.m. energies, together with the CLEO-c measurements. There is a small enhancement around 4.22~GeV, where many charmonium-like states, such as Y(4220), were observed in other processes~\cite{PRL99-182004, PRL118-092001}. The measured BCS at $\sqrt{s}$ = 4.17 GeV agrees with the CLEO-c result, while being $\sim3\sigma$ higher at 3.77 GeV. Figure~\ref{fig_EFF} shows the EFFs determined in this work, in comparison with the CLEO-c measurements~\cite{PRD96-092004} and the theoretical predictions using the covariant spectator quark model~\cite{PRD-074018}. 
The measured EFFs are consistent with the theoretical prediction within an uncertainty of 1.0$\sigma$.

\begin{figure}[htbp]
\centering
\epsfig{width=0.78\textwidth,clip=true,file=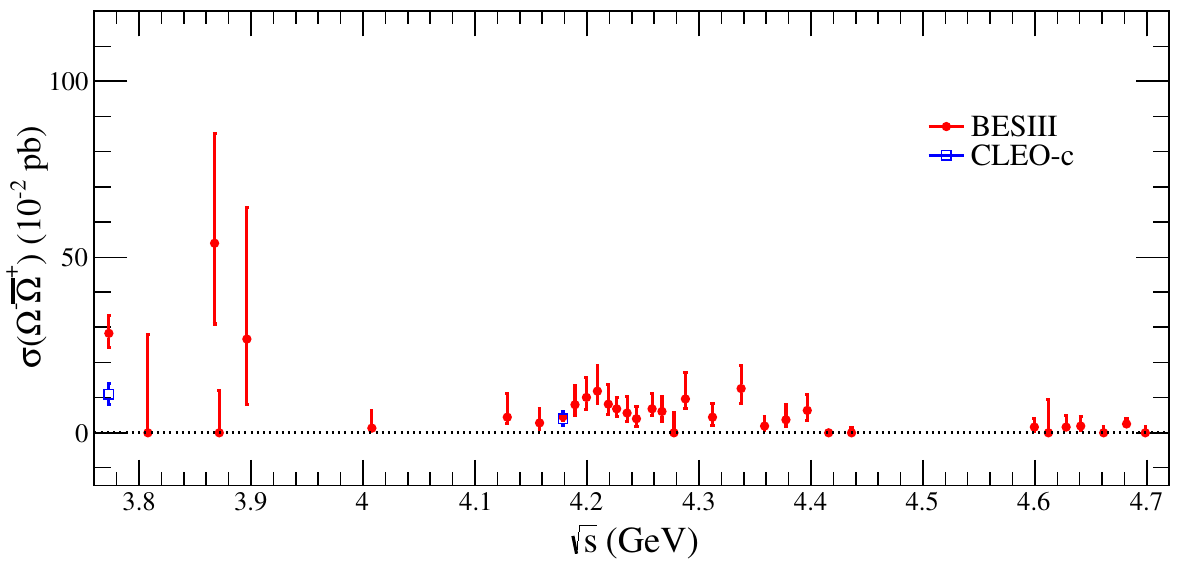}
\caption{The measured $e^+e^-\to \Omega^-\bar{\Omega}^+$ BCSs from 3.7 GeV to 4.7 GeV, where the uncertainties are statistical. The red and blue points with error bars are the results obtained in this work and by CLEO-c~\cite{PRD96-092004}, respectively.}
\label{fig_cros}
\end{figure}

\begin{figure}[htbp]
\centering
\epsfig{width=0.78\textwidth,clip=true,file=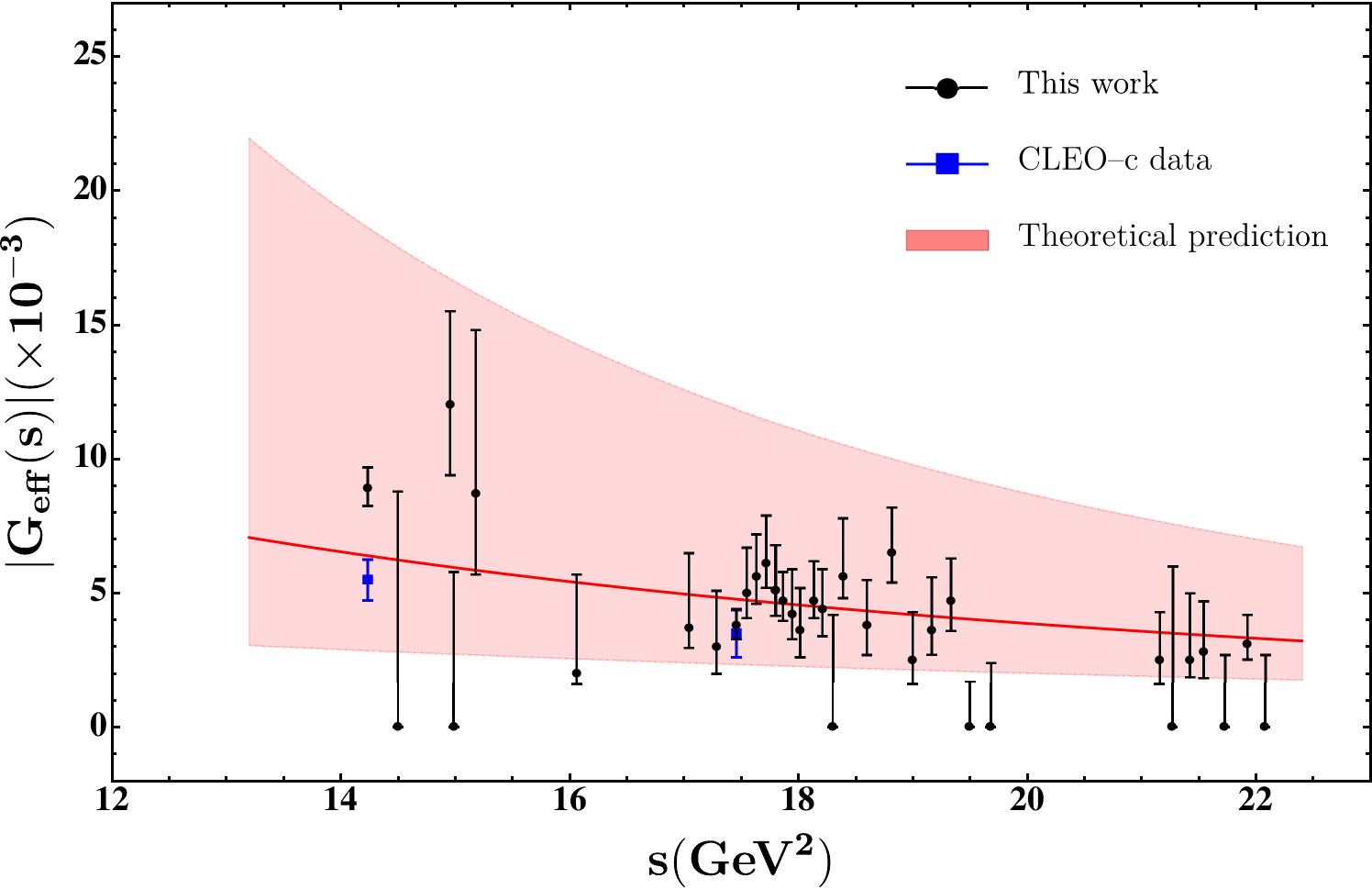}
\caption{Comparison of the measured EFFs and the theoretical prediction. The black points with error bars are the results of this work. The blue points with error bars are the CLEO-c measurements~\cite{PRD96-092004}, where $G_{\text{E0}}$ is assumed to be zero. The red band indicates the theoretical prediction, in which the red line denotes the predicted central value and its boundaries marks the upper and lower limits of the theoretical error~\cite{PRD-074018}.
}
\label{fig_EFF}
\end{figure}

\begin{table}[htbp]
\begin{center}
\caption{The numerical results for $e^+e^-\to \Omega^-\bar{\Omega}^+$
  at each c.m. energy. Here, $\mathcal{L}$ is the integrated
  luminosity~\cite{BAM-340, BAM-522}, $N^{\rm sig}~(N^{\rm UL})$
  denotes the number (upper limit) of the signal events in the signal
  region, 1/$\left|{1-{\Pi}}\right|^2$ is the VP correction factor, $1+\delta$ is
  the ISR correction factor, $\epsilon$ is the signal efficiency $(D$
  = $(1+\delta)\epsilon)$ and $\sigma^{\rm B}$ and $|G_{\rm
    eff}(s)|$ are the measured BCS and EFF, respectively. For the BCS
  and EFF, the first uncertainties are statistical and the second are
  systematic. For $N^{\rm sig}$, $\sigma^{\rm B}$, and $|G_{\rm
    eff}(s)|$, the numbers in brackets correspond to the upper limits
  at the 90\% confidence level. The $S(\sigma)$ is the signal significance estimated by $N^{sig}$/$\sqrt{N^{\rm obs}}$.}

\resizebox{\textwidth}{!}{
\begin{tabular}{cccccccc}
\hline
\hline
$\sqrt{s}$ (GeV)  &  $\mathcal{L}$ (pb$^{-1}$) &  $N^{\rm sig}~(N^{\rm UL})$  & 1/$\left|{1-{\Pi}}\right|^2$  & $D$ (10$^{-2}$) &  $\sigma^{\rm B}$ (fb) & $|G_{\rm eff}(s)|$ (10$^{-3}$) & $S(\sigma)$ \\
\hline

3.7730   & 2931.8  &   50.5$^{+9.0}_{-7.3}$   &1.073& 13.1  &  282.9$^{+50.4}_{-40.9}\pm17.5$ &  8.9$^{+0.8}_{-0.6}\pm0.3$ & 6.2\\
3.8076   & 50.5   &  0.0$^{+0.9}_{-0.0}~(< 2.0)$      &1.071& 13.8 & 0.0$^{+278.7}_{-0.0}\pm$0.0$~(< 619.3)$  & 0.0$^{+8.8}_{-0.0}\pm$0.0$~(< 13.0)$ &...\\
3.8674   & 108.9  &  4.0$^{+2.3}_{-1.7}~(< 8.3)$      &1.056&  14.9  &  539.6$^{+310.3}_{-229.3}\pm$33.5$~(< 1119.7)$ & 12.0$^{+3.5}_{-2.6}\pm$0.4$~(< 18.0)$  &2.0\\  
3.8713   & 110.3  &  0.0$^{+0.9}_{-0.0}~(< 2.0)$       &1.055&  14.8  &  0.0$^{+121.0}_{-0.0}\pm$0.0$~(< 268.8)$ & 0.0$^{+5.8}_{-0.0}\pm$0.0$~(< 8.7)$  &...\\  
3.8962   & 52.6   &  1.0$^{+1.4}_{-0.7}~(< 3.7)$       &1.050& 15.7 &  267.0$^{+373.7}_{-186.9}\pm$16.6$~(< 987.7)$  & 8.7$^{+6.1}_{-3.0}\pm$0.3$~(< 17.0)$ &1.0\\ 
4.0076   & 482.0   &  0.5$^{+1.9}_{-0.2}~(< 3.7)$     &1.036&  17.5&  13.2$^{+50.3}_{-5.3}\pm$0.8$~(< 97.9)$  & 2.0$^{+3.7}_{-0.4}\pm$0.1$~(< 5.3)$  &0.5\\    
4.1285   & 401.5   &  1.5$^{+2.3}_{-0.6}~(< 5.3)$       &1.058&  18.4  &  44.3$^{+67.9}_{-17.7}\pm$2.7$~(< 156.5)$  & 3.7$^{+2.8}_{-0.7}\pm$0.1$~(< 6.9)$ & 1.1\\ 
4.1574   & 408.7   &  1.0$^{+1.4}_{-0.7}~(< 3.7)$       &1.060&  18.6 & 28.7$^{+40.1}_{-20.1}\pm$1.8$~(< 106.0)$  & 3.0$^{+2.1}_{-1.0}\pm$0.1$~(< 5.7)$ &1.0 \\
4.1784   & 3189.0  &  13.0$^{+4.2}_{-3.5}$     &1.062& 19.3 &  46.0$^{+14.9}_{-12.4}\pm2.9$ &  3.8$^{+0.6}_{-0.5}\pm0.1$ & 3.4 \\ 
4.1891   & 526.7   &  3.8$^{+2.6}_{-1.4}~(< 8.3)$         &1.066& 19.5  &  80.1$^{+54.8}_{-29.5}\pm$5.0$~(< 175.1)$ & 5.0$^{+1.7}_{-0.9}\pm$0.2$~(< 7.4)$ & 1.9\\ 
4.1992   & 526.0   &  4.8$^{+2.8}_{-1.7}~(< 9.7)$        &1.068&  19.7 &  100.3$^{+58.5}_{-35.5}\pm$6.2$~(< 202.7)$  & 5.6$^{+1.6}_{-1.0}\pm$0.2$~(< 8.0)$  &2.1\\ 
4.2094   & 517.1   &  5.5$^{+3.3}_{-1.6}~(< 11.1)$        &1.068&  19.3 &  119.1$^{+71.5}_{-34.7}\pm$7.4$~(< 240.4)$ & 6.1$^{+1.8}_{-0.9}\pm$0.2$~(< 8.7)$  &2.2\\ 
4.2189   & 514.6   &  3.8$^{+2.6}_{-1.4}~(< 8.3)$        &1.067&  19.5  &  82.0$^{+56.1}_{-30.2}\pm$5.1$~(< 179.1)$  & 5.1$^{+1.7}_{-0.9}\pm$0.2$~(< 7.5)$  &1.9\\ 
4.2263   & 1056.4   &  6.8$^{+3.2}_{-2.1}~(< 12.4)$     &1.067&  20.4  &  68.4$^{+32.2}_{-21.1}\pm$4.2$~(< 124.7)$ & 4.7$^{+1.1}_{-0.7}\pm$0.1$~(< 6.3)$  &2.6 \\ 
4.2358   & 530.3   &  2.8$^{+2.3}_{-1.2}~(< 6.9)$       &1.064& 20.2 &  56.6$^{+46.5}_{-24.2}\pm$3.5$~(< 139.4)$ & 4.2$^{+1.7}_{-0.9}\pm$0.1$~(< 6.7)$  &1.6 \\ 
4.2440   & 538.1  & 2.0$^{+1.8}_{-1.1}~(< 5.3)$       &1.065&  20.3  & 39.7$^{+35.8}_{-21.9}\pm$2.5$~(< 105.3)$ & 3.6$^{+1.6}_{-1.0}\pm$0.1$~(< 5.8)$ & 1.4\\ 
4.2580   & 828.4   &  5.3$^{+3.5}_{-1.4}~(< 11.1)$        &1.059& 20.5  &  68.0$^{+44.9}_{-18.0}\pm$4.2$~(< 142.5)$ & 4.7$^{+1.5}_{-0.6}\pm$0.1$~(< 6.8)$  &2.2\\ 
4.2668   & 531.1  &  3.0$^{+2.1}_{-1.4}~(< 6.9)$      &1.058& 20.4  &  60.4$^{+42.3}_{-28.2}\pm$3.7$~(< 138.9)$& 4.4$^{+1.5}_{-1.0}\pm$0.1$~(< 6.7)$  &1.7\\
4.2778   & 175.7   &  0.0$^{+0.9}_{-0.0}~(< 2.0)$        &1.057& 20.2 & 0.0$^{+55.5}_{-0.0}\pm$0.0$~(< 123.3)$ & 0.0$^{+4.2}_{-0.0}\pm$0.0$~(< 6.3)$  &...\\
4.2879   & 502.4  &  4.3$^{+3.3}_{-1.2}~(< 9.7)$       &1.056&  19.3 &  96.7$^{+74.2}_{-27.0}\pm$6.0$~(< 218.1)$  & 5.6$^{+2.2}_{-0.8}\pm$0.2$~(< 8.4)$  &1.9\\  
4.3121   & 501.2   &  2.0$^{+1.8}_{-1.1}~(< 5.3)$    &1.055& 19.6 &44.6$^{+40.2}_{-24.6}\pm$2.8$~(< 118.3)$ & 3.8$^{+1.7}_{-1.1}\pm$0.1$~(< 6.2)$  & 1.4\\ 
4.3374   & 505.0   &  5.8$^{+3.0}_{-1.9}~(< 11.1)$       &1.051&  20.0 &  125.9$^{+65.1}_{-41.2}\pm$7.8$~(< 241.0)$ & 6.5$^{+1.7}_{-1.1}\pm$0.2$~(< 9.0)$  &2.4 \\    
4.3583   & 544.0  &  1.0$^{+1.4}_{-0.7}~(< 3.7)$       &1.052&  21.3&  18.9$^{+26.5}_{-13.2}\pm$1.2$~(< 69.9)$  & 2.5$^{+1.8}_{-0.9}\pm$0.1$~(< 4.8)$  &1.0\\ 
4.3774   & 522.7   &  1.8$^{+2.0}_{-0.9}~(< 5.3)$     &1.052& 20.1  &  37.5$^{+41.7}_{-18.8}\pm$2.3$~(< 110.5)$  & 3.6$^{+2.0}_{-0.9}\pm$0.1$~(< 6.1)$  &1.3\\
4.3965   & 507.8  &  3.0$^{+2.1}_{-1.4}~(< 6.9)$       &1.052&  20.2  &  64.3$^{+45.0}_{-30.0}\pm$4.0$~(< 147.9)$ & 4.7$^{+1.6}_{-1.1}\pm$0.1$~(< 7.1)$   &1.7\\ 
4.4156   & 1043.9   &  0.0$^{+0.9}_{-0.0}~(< 2.0)$  &1.055& 21.6  &  0.0$^{+8.7}_{-0.0}\pm$0.0$~(< 19.4)$ & 0.0$^{+1.7}_{-0.0}\pm$0.0$~(< 2.6)$  &...\\ 
4.4362   & 569.9   &  0.0$^{+0.9}_{-0.0}~(< 2.0)$      &1.059&  20.4  &  0.0$^{+16.9}_{-0.0}\pm$0.0$~(< 37.4)$  & 0.0$^{+2.4}_{-0.0}\pm$0.0$~(< 3.6)$  &...\\ 
4.5995   & 586.9   &  1.0$^{+1.4}_{-0.7}~(< 3.7)$      &1.059&  22.4  &  16.6$^{+23.2}_{-11.6}\pm$1.0$~(< 61.4)$  & 2.5$^{+1.8}_{-0.9}\pm$0.1$~(< 4.8)$  &1.0\\ 
4.6119   & 103.8   &  0.0$^{+0.9}_{-0.0}~(< 2.0)$       &1.059&  20.1  &  0.0$^{+94.3}_{-0.0}\pm$0.0$~(< 209.5)$  & 0.0$^{+6.0}_{-0.0}\pm$0.0$~(< 8.9)$ &...\\ 
4.6280   & 521.5  &  0.8$^{+1.6}_{-0.4}~(< 3.7)$        &1.058&  20.1  &  16.6$^{+33.3}_{-8.3}\pm$1.0$~(< 76.9)$  & 2.5$^{+2.5}_{-0.6}\pm$0.1$~(< 5.4)$ & 0.8\\ 
4.6409   & 552.4   &  1.0$^{+1.4}_{-0.7}~(< 3.7)$         &1.058&  20.1  &  19.6$^{+27.5}_{-13.7}\pm$1.2$~(< 72.6)$  & 2.8$^{+1.9}_{-1.0}\pm$0.1$~(< 5.3)$  &1.0\\ 
4.6612   & 529.6  & 0.0$^{+0.9}_{-0.0}~(< 2.0)$        &1.058&  20.1 &  0.0$^{+18.4}_{-0.0}\pm$0.0$~(<  40.9)$  & 0.0$^{+2.7}_{-0.0}\pm$0.0$~(< 4.0)$ &...\\ 
4.6819   & 1669.3   & 3.8$^{+2.6}_{-1.4}~(< 8.3)$         &1.058&  20.0 &  24.8$^{+17.0}_{-9.1}\pm$1.5$~(< 54.2)$ & 3.1$^{+1.1}_{-0.6}\pm$0.1$~(< 4.6)$  &1.9 \\
4.6988   & 536.5   &  0.0$^{+0.9}_{-0.0}~(< 2.0)$      &1.058& 20.0  &  0.0$^{+18.3}_{-0.0}\pm$0.0$~(< 40.7)$   & 0.0$^{+2.7}_{-0.0}\pm$0.0$~(< 4.0)$ &...\\

\hline
\hline
\end{tabular}
}
\label{cros_list}
\end{center}
\end{table}

\section{Systematic uncertainty}
\label{chap:SYSTEMATICS}

The systematic uncertainties in the BCS measurement mainly come from
the luminosity measurement,  branching fractions of the
$\Omega^-\to K^-\Lambda$ and $\Lambda\to p\pi^-$ decays, 
$\Omega^{-}$ reconstruction efficiency,  MC generator,  mass
windows of the $\Lambda$ and $\Omega^-$ selections,  choice of the
sideband regions and  $(1+\delta)\epsilon$ estimation. All
sources of systematic uncertainties are
discussed in detail below.

The integrated luminosity is measured using Bhabha events with an uncertainty of 1.0\%~\cite{CPC39-093001}.
The uncertainties of the branching fractions of $\Omega^-\to K^-\Lambda$ and $\Lambda\to p\pi^-$ are taken from the PDG~\cite{pdg}.
The systematic uncertainty due to the $\Omega$ reconstruction
efficiency, including the tracking and PID efficiencies of charged
tracks and the efficiency of the $\Lambda$ reconstruction, is
estimated by the control sample
$\psi^{\prime}\to\Omega^{-}\bar{\Omega}^+$ with the same method as
described in Ref.~\cite{PRD104-L091104}. We consider the systematic
uncertainties of the decay parameters ($\alpha_{\Omega}$ and
$\alpha_{\Lambda}$) in the angular distribution function and the
helicity parameters from the analysis of $\psi^{\prime}\to
\Omega^{-}\bar{\Omega}^+$ events~\cite{songjiaojiao} as the systematic
uncertainty for the MC generator.

The uncertainties of the $M(p\pi^-)$ and $M^{\rm cor}(K^-\Lambda)$ mass windows are estimated by varying their sizes by $\pm$1$\sigma$. The systematic uncertainty of the $RM^{\rm cor}(K^-\Lambda)$ mass window is estimated by varying it by $\pm$0.01~GeV$/c^{2}$. The larger deviation in each case is taken as the uncertainty. To estimate the uncertainty from the sideband positions, we double the relative distance between the signal and sideband regions. After adjusting the box window, we compare the signal yield from data and the efficiency from MC simulation; the relative difference between the signal yield ratio from data and the efficiency ratio from MC is taken as the systematic uncertainty~\cite{BESIII:2012ghz,mass window}. 
The $(1+\delta)\epsilon$ systematic uncertainty depends on two
different sources. The input line-shape of the cross section
is iterated until the final cross section becomes stable, and the
largest variation in $(1+\delta)\epsilon$ (1.8\%) during the last
two iterations is taken as the systematic uncertainty.
The ISR factor and the signal efficiency are dependent on the fit to
the line shape of the cross section. The systematic uncertainty is evaluated by varying the fitted line-shape parameters by +$1\sigma$ of statistical uncertainty in the input cross-section model. The resulting largest difference of
$(1+\delta)\epsilon$ (2.0\%) is assigned as the systematic uncertainty.

All systematic uncertainties discussed above are summarized in Table~\ref{tab:sys}.  Assuming all sources are independent, the total systematic uncertainty in the BCS measurement is determined to be 6.9\% by adding them in quadrature. 

\begin{table}[htbp]
\begin{center}
\caption{Systematic uncertainties in the BCS measurement for all data samples~(in \%). The uncertainties are the same for the thirty-four c.m. energy points.}
\begin{tabular}{lc}
\hline
Source  & Uncertainty\\  
\hline
\hline
Luminosity                       &     1.0  \\
$\mathcal{B}(\Omega^-\to K^-\Lambda)$   & 0.7\\
$\mathcal{B}(\Lambda\to p\pi^-)$        &  0.5  \\
$\Omega^-$ reconstruction          &  0.7   \\
MC generator                 &  2.2  \\
$M(p\pi^-)$ mass window    &  2.6  \\
$M^{\rm cor}(K^-\Lambda)$ mass window     &    2.6 \\
$RM^{\rm cor}(K^-\Lambda)$ mass window     &    3.5 \\
Position of the sidebands &     2.7\\
$(1+\delta)\epsilon$        &    2.7  \\
\hline                                      
Sum in quadrature                &  6.9   \\
\hline
\hline
\end{tabular}
\label{tab:sys}
\end{center}
\end{table}

\section{Summary}
\label{chap:SUMMARY}

In summary, using 22.7 fb$^{-1}$ of $e^+e^-$ collision data collected
at thirty-four c.m. energies from 3.7 GeV to 4.7 GeV with the BESIII
detector, the BCSs of the process $e^+e^-\to \Omega^-\bar{\Omega}^{+}$ and the EFFs
of $\Omega^-$ are determined by the single baryon tag method. Clear $e^+e^-\to
\Omega^-\bar{\Omega}^{+}$ signals are observed at $\sqrt{s}$ = 3.77 and 4.17~GeV. From the $\sqrt{s}$-dependent BCS, no significant resonance contribution is observed. A hint of enhancement around 4.2 GeV is observed, but we can't draw clear conclusion due to large uncertainty. More data is needed to confirm  the structure in this region. The experimental measurements of the EFFs are consistent with the
theoretical calculation. These results obtained provide important
information to understand the production mechanism of $\Omega^-\bar{\Omega}^{+}$ pair and the inner structure
of the $\Omega^-$ baryons, which are crucial to constrain the
theoretical predictions~\cite{PRD-074018, Ramalho:2019koj}.



\acknowledgments

The BESIII Collaboration thanks the staff of BEPCII (https://cstr.cn/31109.02.BEPC) and the IHEP computing center for their strong support. This work is supported in part by National Key R\&D Program of China under Contracts Nos. 2023YFA1606704, 2023YFA1606000; National Natural Science Foundation of China (NSFC) under Contracts Nos. 11635010, 11935015, 11935016, 11935018, 12025502, 12035009, 12035013, 12061131003, 12192260, 12192261, 12192262, 12192263, 12192264, 12192265, 12221005, 12225509, 12235017, 12361141819; the Chinese Academy of Sciences (CAS) Large-Scale Scientific Facility Program; CAS under Contract No. YSBR-101; 100 Talents Program of CAS; The Institute of Nuclear and Particle Physics (INPAC) and Shanghai Key Laboratory for Particle Physics and Cosmology; ERC under Contract No. 758462; German Research Foundation DFG under Contract No. FOR5327; Istituto Nazionale di Fisica Nucleare, Italy; Knut and Alice Wallenberg Foundation under Contracts Nos. 2021.0174, 2021.0299; Ministry of Development of Turkey under Contract No. DPT2006K-120470; National Research Foundation of Korea under Contract No. NRF-2022R1A2C1092335; National Science and Technology fund of Mongolia; Polish National Science Centre under Contract No. 2024/53/B/ST2/00975; STFC (United Kingdom); Swedish Research Council under Contract No. 2019.04595; U. S. Department of Energy under Contract No. DE-FG02-05ER41374.


\bibliographystyle{JHEP}

\end{document}